\documentclass[twocolumn]{aastex701}

\usepackage{amsmath}
\usepackage{bm}

\begin{document}

\title{Physics Informed Neural Network for Multi-dimensional Turbulent Self-gravitating Flows}

\author[0000-0002-0164-2721]{Tirthankar De}
\affiliation{Indian Institute of Technology Roorkee, Roorkee - Haridwar Highway, Roorkee, Uttarakhand, India - 247667}
\email{tirthankar_d@ph.iitr.ac.in}  

\author[0000-0003-3784-8913]{Sayantan Auddy}
\affiliation{Fields Institute for Research in Mathematical Sciences, 222 College Street, Toronto, Ontario M5T 3J1,
Canada}
\email{sayantanauddy21@gmail.com}

\author{Ramit Dey}
\affiliation{Fields Institute for Research in Mathematical Sciences, 222 College Street, Toronto, Ontario M5T 3J1,
Canada}
\email{ramitdey@gmail.com}

\author[0000-0003-0855-350X]{Shantanu Basu}
\affiliation{Canadian Institute for Theoretical Astrophysics, University of Toronto, 60 St. George Street, Toronto, ON M5S 3H8, Canada}
\affiliation{Department of Physics and Astronomy,
University of Western Ontario,
London, Ontario N6A 3K7, Canada}
\email{basu@uwo.ca}

\begin{abstract}

Molecular clouds exhibit complex, multi-scale dynamics driven by the interplay of turbulence, self-gravity, and gas pressure. Modelling these systems with conventional grid-based methods becomes increasingly expensive as spatial resolution and dimensionality increase. Physics-Informed Neural Networks (PINNs) offer a mesh-free alternative by incorporating the governing equations directly into the training objective, but their ability to model realistic turbulent self-gravitating flows remains largely unexplored. We apply PINNs to isothermal molecular clouds initialized with astrophysically motivated turbulent velocity fields, considering subsonic and supersonic evolution in two and three dimensions. We benchmark the predicted density and velocity fields against the finite-volume code Athena++. PINNs accurately recover the global evolution of the flow over multiple free-fall times across turbulent realizations, Mach numbers, and dimensionalities. Their training cost also exhibits weak dependence on spatial resolution, increasing by less than $25\%$ from $128^3$ to $1024^3$ in our experiments. These results demonstrate that PINNs can model substantially more realistic self-gravitating turbulent flows than previously explored and provide a promising complementary framework for studying these systems.

\end{abstract}

\keywords{\uat{Hydrodynamics}{1963} --- \uat{Star formation}{1569} --- \uat{Gravitational Instability}{668} --- \uat{Neural Networks}{1933}}


\section{Introduction}

Molecular clouds are the sites of star formation and are characterized by the interplay of self-gravity, gas pressure, and turbulent motions. Highly structured, multi-scale gas flows shaped by the interplay between self-gravity and turbulence exist in these systems. Observations reveal substantial velocity structure across a broad range of spatial scales, while numerical studies show that supersonic turbulence can strongly influence the formation and
evolution of dense structures 
\citep{heyer04, mckee07}.
Capturing the coupled evolution of density, velocity, and gravitational potential in such flows therefore provides a stringent test for numerical methods, particularly when the dynamics become strongly nonlinear and span multiple spatial scales.

Grid-based methods provide a well-established framework for solving self-gravitating hydrodynamics and have been successfully applied to turbulent molecular-cloud simulations. However, resolving a broad range of spatial scales can require increasingly fine computational meshes, with the associated cost becoming particularly significant in three dimensions. These considerations motivate the exploration of alternative numerical formulations that do not rely on a fixed spatial discretization.

Physics-informed neural networks (PINNs) provide one such formulation \citep{raissi2019physics}. Instead of representing the solution on a prescribed mesh, a neural network can represent the physical fields as continuous functions of space and time,
while the governing differential equations are incorporated directly into the training objective through automatic differentiation. This approach allows the equations themselves to provide the physical constraints during training and does not require a labelled solution dataset. PINNs have therefore attracted increasing interest in astrophysical applications \citep{mao2025neural}, although their ability to represent strongly nonlinear, turbulent, self-gravitating flows remains comparatively unexplored.

The application of PINNs to self-gravitating hydrodynamics was introduced in the Gravity-Informed Neural Network (GRINN) framework of \citet{auddy2024grinn}. GRINN demonstrated that a physics-informed neural network could reproduce the evolution of isothermal self-gravitating gas in up to three dimensions. In that work, the initial conditions consisted of
controlled sinusoidal perturbations imposed on an otherwise uniform background. These initial conditions provided a useful setting for testing the method against analytic linear solutions and conventional numerical calculations, and established a baseline for applying PINNs to self-gravitating astrophysical flows. However, a sinusoidal perturbation
primarily probes a prescribed spatial mode and does not capture the broadband, stochastic structure characteristic of turbulent molecular-cloud flows.

In this work, we extend the GRINN framework from controlled perturbations to broadband turbulent initial conditions. We consider an idealized, isothermal, self-gravitating gas with turbulent velocity fields containing power across a range of spatial scales, and compare the resulting PINN solutions with calculations performed using \texttt{Athena++}
\citep{Stone2020}. We examine subsonic and supersonic flows in two dimensions and subsonic flows in three dimensions, while also testing multiple turbulent realizations and reference-grid resolutions. This provides a more stringent test of whether a relatively simple physics-informed formulation can retain predictive capability when the
initial conditions contain multi-scale structure and the subsequent evolution becomes nonlinear.

We find that the PINN reproduces the dominant density structure and large-scale flow organization across a substantial portion of the parameter space, including turbulent flows in both two and three dimensions. At fixed neural network architecture, optimization budget, and collocation-point budget, the PINN training cost also shows comparatively weak dependence on the resolution of the reference grid. The most strongly supersonic three-
dimensional case, however, exposes a clear limitation of the present formulation, with substantially larger discrepancies from the reference solution during gravitational collapse. Rather than replacing established grid-based methods, these results demonstrate that physics-informed neural networks can capture meaningful aspects of complex self-gravitating astrophysical flows while also revealing specific regimes in which more expressive architectures and improved training strategies are needed. We view this as a step toward developing physics-informed machine-learning methods capable of incorporating increasingly realistic astrophysical physics.

\begin{figure*}
  \centering
  \includegraphics[width=\textwidth]{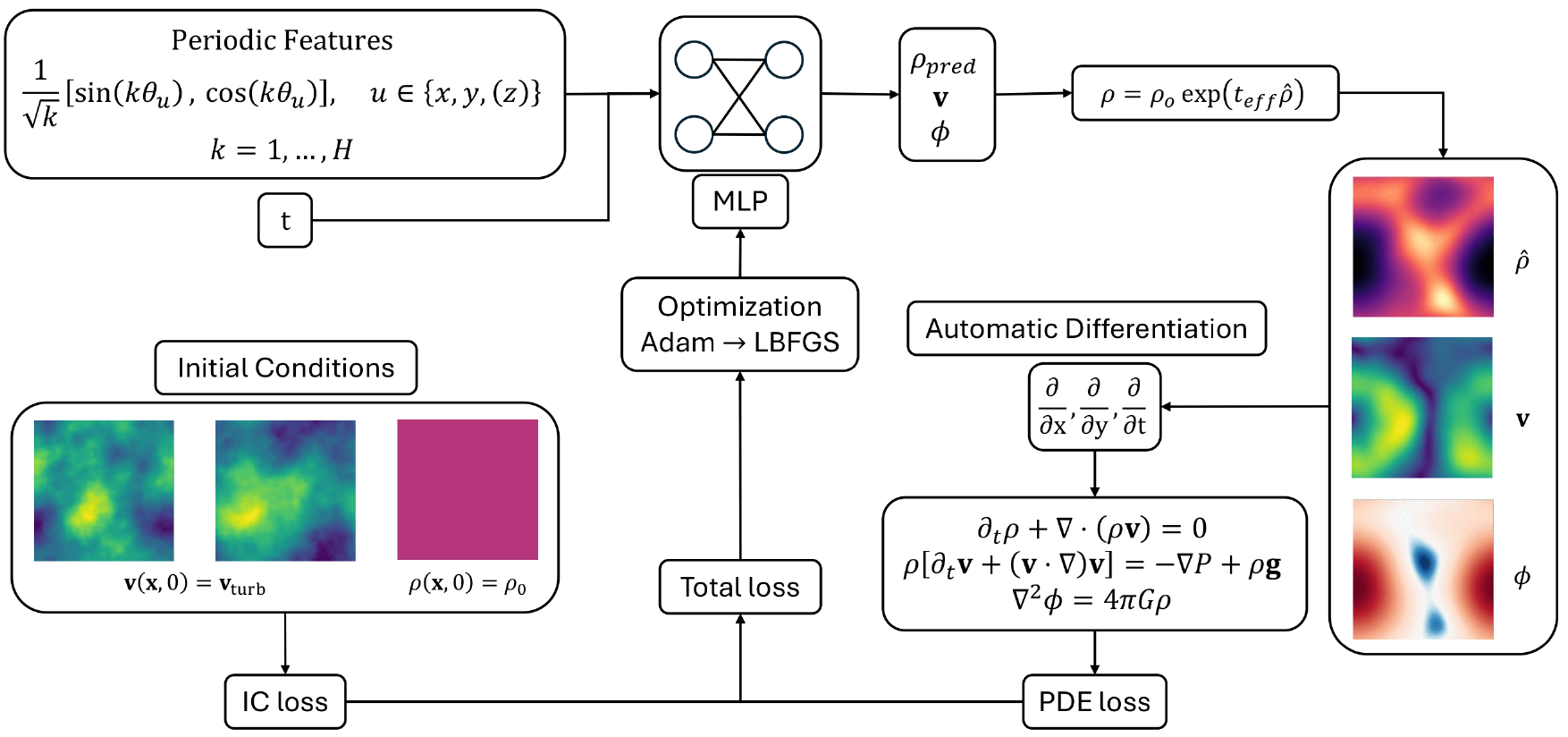}
  \caption{Schematic of the PINN training pipeline. Spatial coordinates are mapped to periodic features while time is passed to the network unchanged; the MLP then predicts the transformed density field, velocity, and gravitational potential (Section~\ref{sec:pinn}; Appendices~\ref{app:fourier} and~\ref{app:exparam}). Automatic differentiation evaluates the governing-equation residuals, while the initial-condition and PDE residuals define the training loss (Appendix~\ref{app:loss}). Training proceeds sequentially with Adam and L-BFGS. The initial-condition maps show the turbulent velocity components (for 2D cases) and the initially uniform density; the initial gravitational potential is zero everywhere (Appendix~\ref{app:icgen}).}
  \label{fig:Architecture}
\end{figure*}

\section{Physical model}

\subsection{Hydrodynamic system}

We consider an isothermal, self-gravitating gas, an approximation commonly adopted for cold molecular-cloud gas, when radiative cooling is sufficiently efficient to maintain approximately constant temperature \citep{banerjee2004formation, chevance2020molecular}. The gas dynamics are governed by the continuity, momentum, and Poisson equations,

\begin{equation}
\partial_t \rho + \bm{\nabla} \cdot (\rho\, \bm{v}) = 0\, , \label{eq:continuity}
\end{equation}
\begin{equation}
\rho\left[\partial_t \bm{v} + (\bm{v} \cdot \bm{\nabla})\bm{v}\right]
    = -\bm{\nabla} P + \rho\, \bm{g}\, , \label{eq:momentum}
\end{equation}
\begin{equation}
\bm{\nabla}^2 \phi = 4\pi G\rho, \label{eq:poisson}
\end{equation}
where $\bm{g} = -\bm{\nabla}\phi$ is the gravitational field, $G$ is the gravitational constant, and the pressure closes the system via the isothermal relation $P = c_s^2\rho$, where $c_s = (kT/m)^{1/2}$ is the isothermal sound speed.  We impose periodic boundary conditions throughout.

We adopt the set of units in \citet{auddy2024grinn}, setting
$c_s=\rho_0=4\pi G=1$. The corresponding characteristic units of length
and time are
\begin{equation}
L_0=\frac{c_s}{\sqrt{4\pi G\rho_0}},
\qquad
t_0=\frac{1}{\sqrt{4\pi G\rho_0}}.
\end{equation}
All lengths and times quoted below are in units of $L_0$ and $t_0$, respectively. With this normalization, the thermal Jeans wavenumber is $k_J=1$, giving a Jeans wavelength
$\lambda_J=2\pi L_0$. Our computational domain has a side length $L=14L_0$, corresponding to $L/\lambda_J\simeq2.2$ and hence containing gravitationally unstable modes. Unlike the single-mode perturbation used in \citep{auddy2024grinn}, our turbulent initial conditions are broadband, with a velocity power spectrum $P_v(k)\propto k^{-4}$, so no single
perturbation wavelength characterizes the initial velocity field.

\subsection{Initial conditions}

The gas is initialized with uniform density $\rho(\bm{x},0)=\rho_0$ and zero gravitational potential $\Phi(\bm{x},0)=0$.  The velocity field is seeded by a turbulent realization,
\begin{equation}
\bm{v}(\bm{x},0) = \bm{v}_{\rm turb}(\bm{x}),
\end{equation}
generated from a Gaussian random field with a prescribed Fourier-space power
spectrum
\begin{equation}
P_v(k) \propto k^{-4},
\end{equation}
characteristic of highly compressible, shock-dominated turbulence that is widely adopted in
molecular-cloud simulations \citep{auddy2018effect, mac2004control}.
The field is synthesised by drawing independent white-noise realisations for each
velocity component \citep{federrath2010comparing}, filtering them in Fourier space with the power-law amplitude
$|\tilde{v}(\bm{k})| \propto k^{-2}$ (so that $P_v \propto |\tilde{v}|^2 \propto k^{-4}$),
and transforming back to physical space.
The $\bm{k}=\bm{0}$ mode is set to zero, ensuring each component has zero spatial mean and hence no net bulk flow.
Each component is then rescaled so that its spatial root-mean-square equals $a\,c_s$ exactly, giving the target per-component Mach number. All density structure that develops during the simulation therefore arises self-consistently from turbulent compression and gravitational instability. The same turbulent realisation initialises both the PINN and the Athena++ reference simulation: the velocity field is generated once on a regular Cartesian grid and bilinearly interpolated to both the finite-volume cell centres and the PINN
collocation points, so that differences in the resulting solutions arise solely
from the numerical framework (details in Appendix~\ref{app:icgen}).

\begin{figure*}
  \centering
  \includegraphics[width=\textwidth]{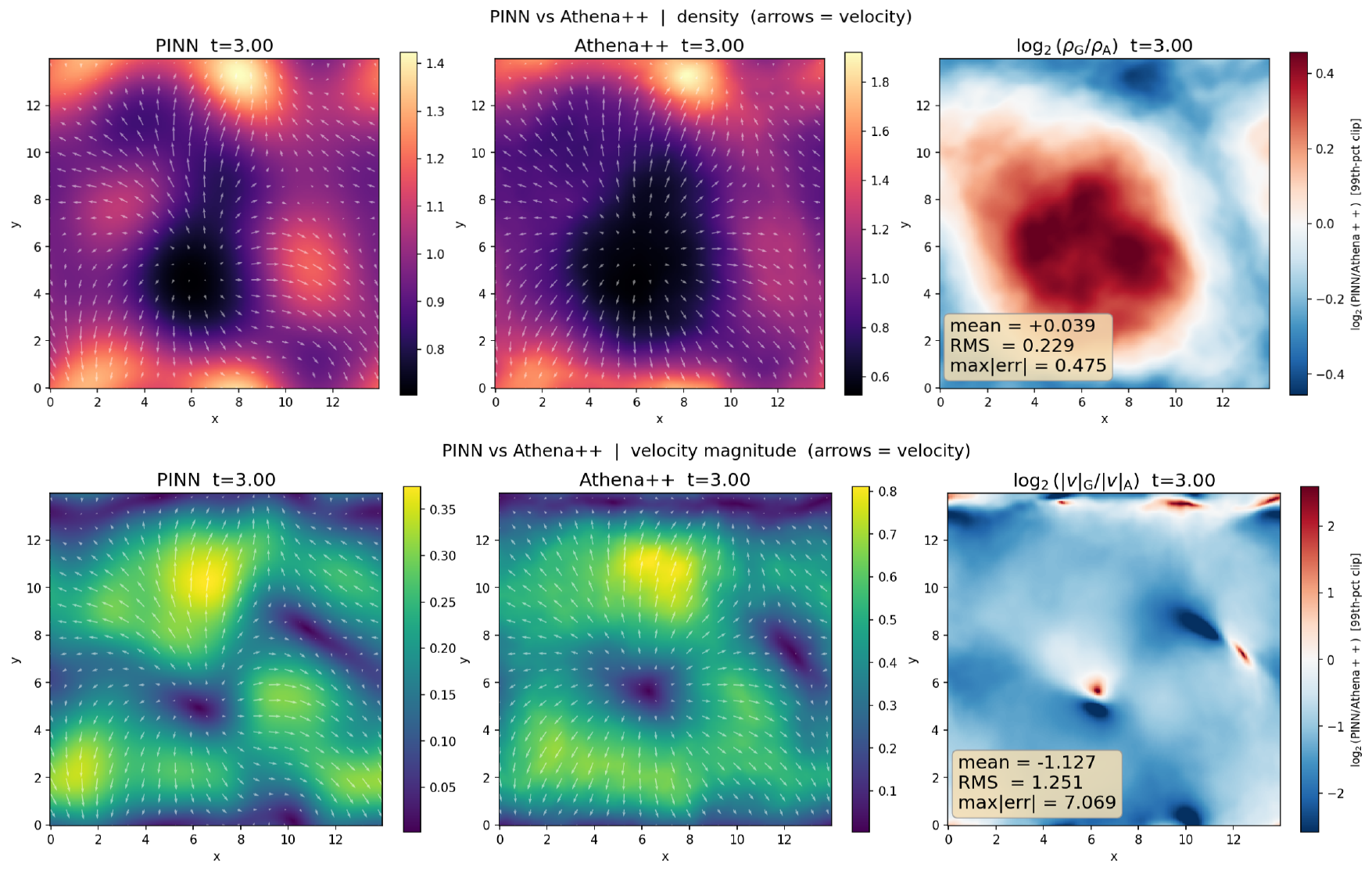}
  \caption{Comparison of the PINN and Athena++ solutions for the 2D subsonic case ($a=0.1$) at $t=3.00\,t_0$. The top row shows the density field, and the bottom row shows the velocity magnitude, $\lvert\bm{v}\rvert=\sqrt{v_x^2+v_y^2}$. In each row, the left and middle panels show the PINN and Athena++ solutions, respectively, while the right panel shows the pixel-wise logarithmic ratio, $\log_2(f_G/f_A)$, where $f$ denotes the density or velocity magnitude, and subscripts $G$ and $A$ denote the PINN and Athena++ solutions, respectively. The annotations report the mean signed log-ratio, its RMS, and the maximum absolute log-ratio; the colour scale is clipped at the 99th percentile of the absolute log-ratio. Arrows indicate the velocity direction.}
  \label{fig:2d_subsonic}
\end{figure*}

The perturbation amplitude $a$ is defined through the one-dimensional RMS velocity dispersion,
\begin{equation}
\sigma_{v_i} = a\,c_s,
\end{equation}
giving a total RMS Mach number $\mathcal{M}_{\rm rms} = \sqrt{d}\,a$, where $d$ is the number of spatial dimensions.  Subsonic cases satisfy $a < 1$ and supersonic cases $a > 1$.

\begin{figure*}
  \centering
  \includegraphics[width=0.99\textwidth]{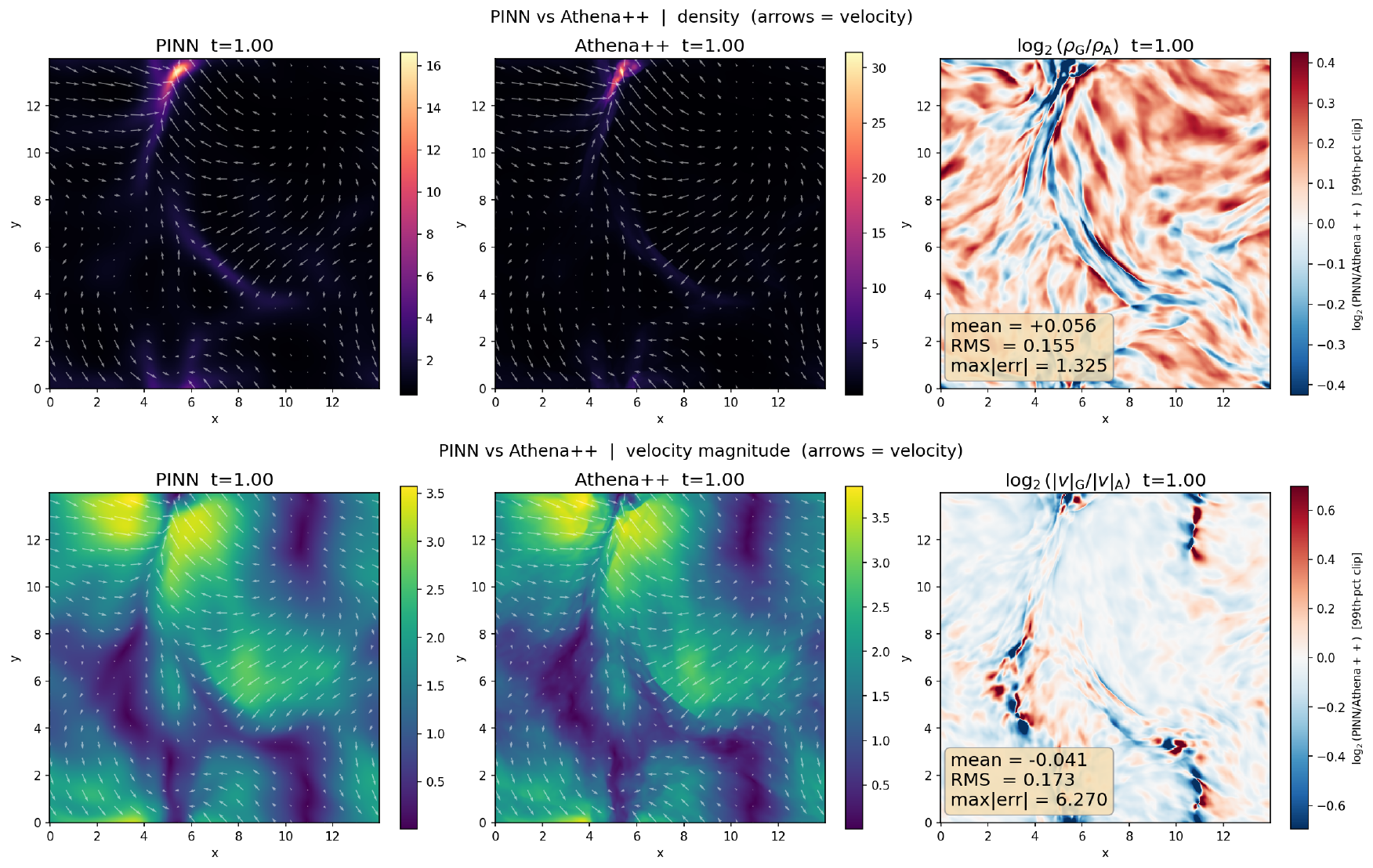}
  \caption{Same as Figure \ref{fig:2d_subsonic}, but with $a = 1.1$, representing a supersonic turbulence in the system.}
  \label{fig:2d_supersonic}
\end{figure*}

\section{PINN}
\label{sec:pinn}

We represent the density, velocity, and gravitational potential of the self-gravitating flow as continuous functions of space and time using a physics-informed neural network (PINN). The network parameters are determined by minimizing the residuals of the governing hydrodynamic and Poisson equations at collocation points distributed throughout the spatio-temporal domain. Because the solution is represented as a continuous function, the trained network can be evaluated at arbitrary spatial and temporal coordinates within the training domain. Automatic differentiation \citep{wengert1964simple} is used to evaluate the derivatives of the network outputs required by the PDE residuals. Unlike data-driven surrogate models, the PINN is trained without labelled solution data: the Athena++ solutions are used only for post-training validation and do not enter the training loss.

The PINN takes as input the spatial coordinates and time. Because the simulation domain is periodic, each spatial coordinate is mapped to periodic Fourier features before being supplied to the network. This representation enforces periodicity by construction, so no explicit boundary-condition loss is required. Details of the Fourier feature construction are given in Appendix~\ref{app:fourier}.

The network predicts the velocity components and gravitational potential, together with a transformed representation of the density. We parametrize the physical density exponentially, such that $\rho>0$ for all network outputs. The same parametrization also enforces the uniform initial density $\rho(\bm{x},0)=\rho_0$ exactly as a hard constraint, rather than requiring the network to learn this condition through a loss term. The prescribed turbulent velocity field is imposed through the initial-condition loss.
Further details of the density parametrization and initial-condition enforcement are given in Appendix~\ref{app:exparam} and Appendix~\ref{app:loss}.

At the start of the calculation, the turbulent velocity perturbation is imposed instantaneously while the initially uniform density has not yet responded to the resulting velocity field. We therefore introduce a short startup interval $\Delta t_s=0.1t_0$ before enforcing the PDE residuals. This allows the initial velocity perturbation to develop a
consistent density response before the full hydrodynamic constraints are applied (Appendix~\ref{app:startup}).

Training proceeds in two sequential phases. In the first phase, the Adam optimizer \citep{zhang2018improved} is first used to bring the network to a converged representation of the flow. In the second phase, the L-BFGS optimizer \citep{liu1989limited} refines the solution to higher precision. We use \textit{sine} activation functions throughout, which provides better gradient flow and training stability than saturating activations such as sigmoid \cite{parascandolo2017taming}. The full loss formulation and implementation details are given in Appendix~\ref{app:loss}.

\begin{figure*}
  \centering
  \includegraphics[width=0.995\textwidth]{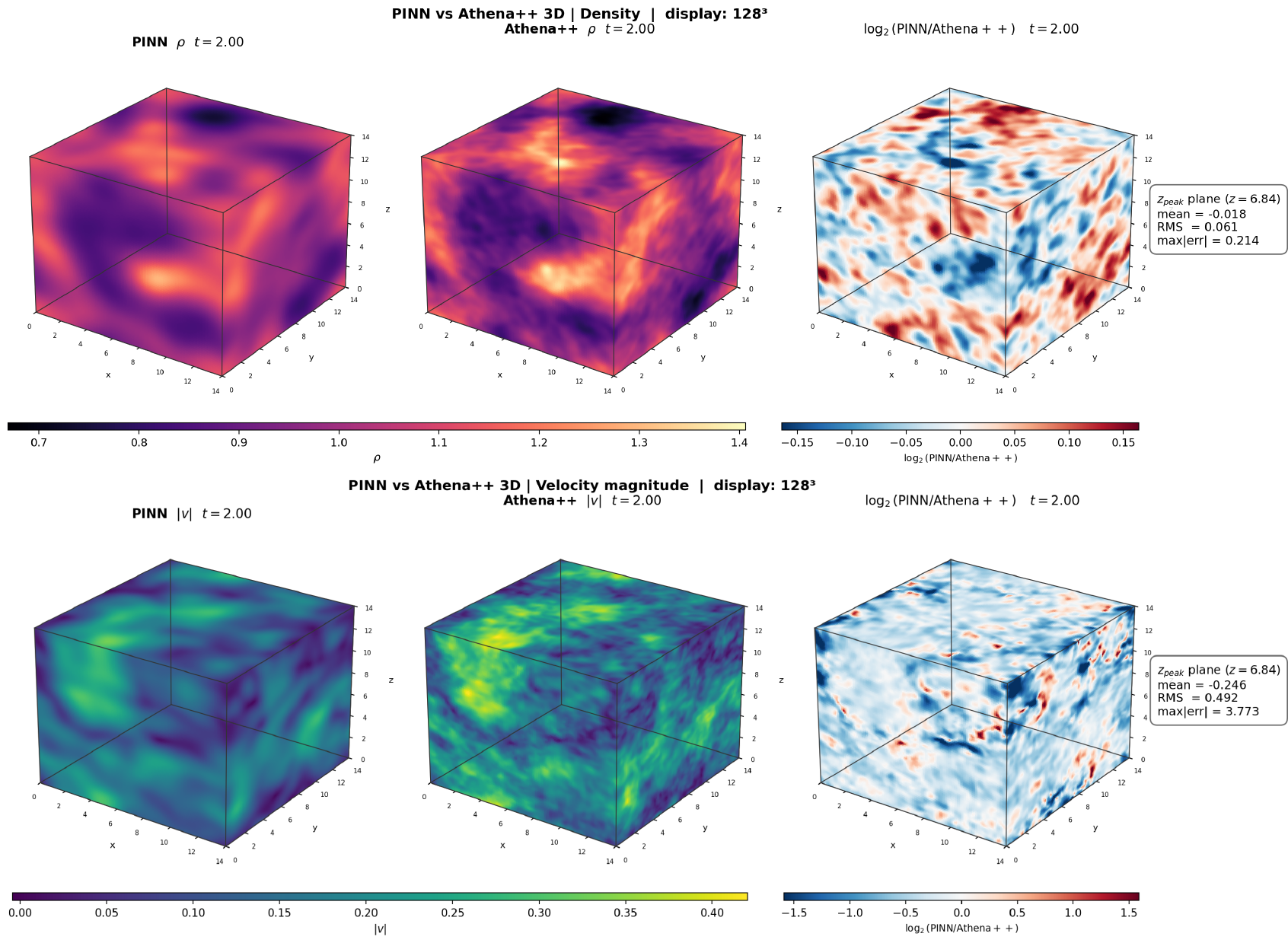}
  \caption{Comparison for PINN vs Athena++ in a 2D subsonic perturbed ($a = 0.1$) system. The display value at the title represents the resolution value of the IC field supplied to the PINN and Athena++ (details in \ref{app:icgen}). The values in the box at the right show the z coordinate value of the collapsing core, while the metrics below show the comparison between PINN and Athena++ in the plane $z = z_{peak}$}
  \label{fig:3d_subsonic}
\end{figure*}

\subsection{Computational implementation}

The PINN training was performed on a single NVIDIA H100 GPU using a 3g.40GB MIG partition. Athena++ was compiled with MPI support and executed using distributed MPI parallelism. The Athena++ MeshBlock size was fixed at $64^3$ cells for all runs. This configuration allows the computational domain to be distributed over an increasing number of MeshBlocks while maintaining a fixed local problem size.

\section{Results}

We assess PINN performance against Athena++ using the pixel-wise signed log-ratio,
\[
\delta = \log_2\!\left(\frac{f_G}{f_A}\right),
\]
where $f$ denotes either the density $\rho$ or the velocity magnitude $\lvert\bm{v}\rvert$, and the subscripts $G$ and $A$ denote the PINN and Athena++ solutions, respectively. A value of $\delta = 0$ indicates perfect agreement; $|\delta| = 1$ corresponds to a factor-of-two discrepancy. Since $\delta$ is logarithmic, an RMS value $\delta_{\rm RMS}$ corresponds to a characteristic multiplicative discrepancy of $2^{\delta_{\rm RMS}}-1$; thus, $\delta_{\rm RMS}=0.15$ corresponds to a characteristic discrepancy of approximately $11\%$. The mean signed bias $\langle\delta\rangle$, its RMS, and the maximum absolute error are annotated on each panel; the colorbar is clipped at the 99th percentile of $|\delta|$. For 3D runs, each panel is a three-dimensional box rendering of the computational domain, showing three
orthogonal faces of the cube: the top, front and right. Beside each bias panel we report the mean, RMS and maximum absolute error evaluated on the XY plane at the instantaneous density maximum, with the PINN evaluated at the native Athena++ grid points on that plane.

\subsection{Case 1: Subsonic perturbations in 2D}

Figure \ref{fig:2d_subsonic} shows the performance of our model for $a = 0.1$. We see that the PINN captures the broad density distribution and associated velocity field with close visual agreement to Athena++ at both early and late times. At $t = 3.00$, the density RMS log-ratio is $0.229$, corresponding to a characteristic multiplicative discrepancy of approximately $17\%$. The errors concentrate around the small-scale, high-gradient structures, while the large-scale density organization remains consistent between the two solutions.

\subsection{Case 2: Supersonic perturbations in 2D}

We next investigate a supersonic case in Figure \ref{fig:2d_supersonic} with $a = 1.1$. The temporal range of comparison is shorter here as gravitational collapse proceeds faster at higher Mach numbers. The PINN is still able to reproduce the main density and velocity morphology seen in Athena++. At $t = 1.00$, the density RMS log-ratio remains $0.155$, corresponding to a characteristic multiplicative discrepancy of approximately $11\%$. The agreement is strongest in the large-scale flow organization, while discrepancies are more visible in localized shocked or filamentary regions, where sharper gradients and smaller-scale structures are more prominent. The error maps show that the remaining differences are not spatially uniform, but instead track the most nonlinear parts of the flow.

\subsection{Case 3: Subsonic perturbations in 3D}

To assess whether the network can learn the dynamics of a 3D system, we compare PINN and Athena++ performance in a similar way as 2D for $a = 0.1$, shown in Figure \ref{fig:3d_subsonic}. For the 3D subsonic case, the density RMS log-ratio is $0.061$, corresponding to a characteristic multiplicative discrepancy of approximately $4.3\%$. The comparison confirms that the PINN generalizes beyond 2D and remains capable of tracking the dominant gravitationally driven evolution in three dimensions. The largest discrepancies again occur in smaller-scale structures, where the PINN solution is smoother than the grid-based result and under-represents some of the finer vortical or compressive features.

Thus across all benchmark cases, the PINN consistently reproduces the global density and velocity structure of the self-gravitating flow, with the remaining differences concentrated in localized fine-scale features.

\subsection{Computational Scaling}
\label{sec:scaling}

\begin{figure*}[t]
  \centering
  \includegraphics[width=0.49\textwidth]{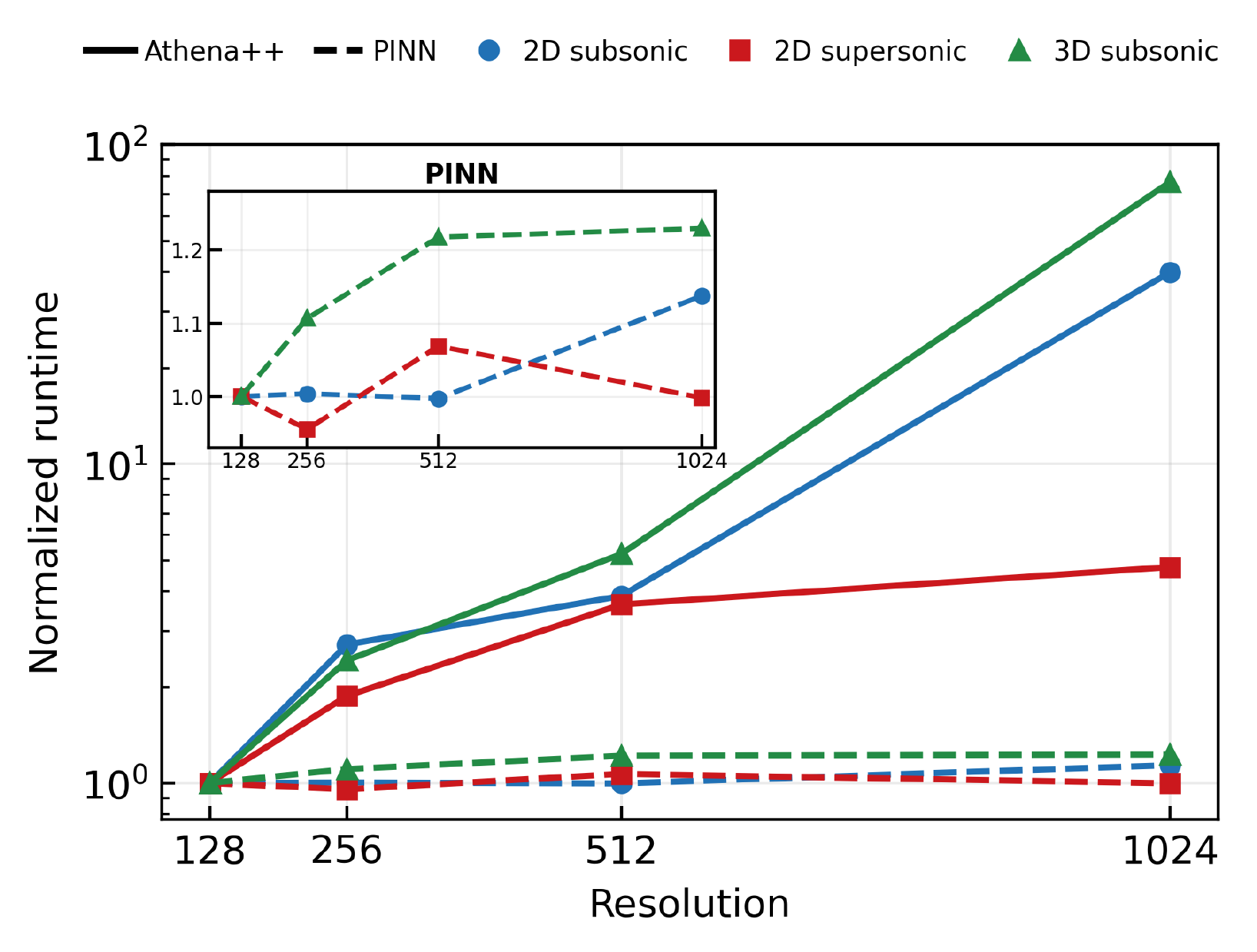}\hfill
  \includegraphics[width=0.49\textwidth]{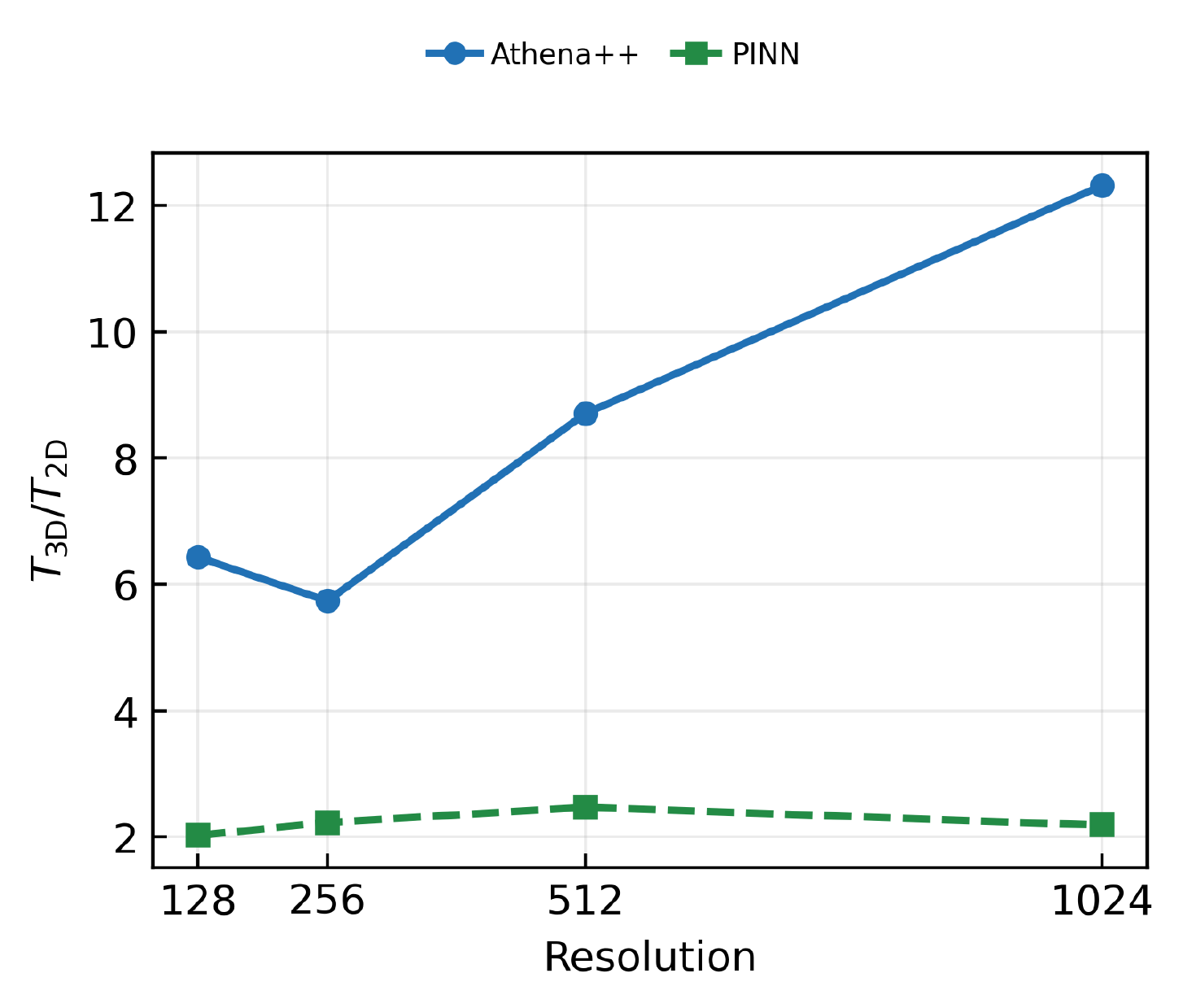}
  \caption{(\textit{Left}) Normalized runtime of PINN and Athena++ as a function of spatial resolution for the 2D subsonic, 2D supersonic, and 3D subsonic cases, each normalised to its own $128$-resolution baseline. The inset provides a magnified view of the PINN curves. (\textit{Right}) Runtime ratio $T_{\rm 3D}/T_{\rm 2D}$ for PINN and Athena++ at each resolution, computed from the subsonic runs. PINN's dimensional overhead remains nearly constant near $2\times$, while Athena++'s grows from ${\approx}6\times$ to ${\approx}12\times$ with increasing resolution.}
  \label{fig:scaling}
\end{figure*}

Figure~\ref{fig:scaling} (left) shows normalized runtime as a function of spatial resolution for all of the above shown three cases, each normalized to its own $128$-resolution baseline. Athena++ costs grow steeply with resolution: the 3D subsonic case requires ${\approx}76\times$ longer at resolution $1024$ than at $128$, while the 2D cases show $40\text{--}76\times$ increases. In contrast, PINN training time increases by less than $25\%$ over the same resolution range.

Because the collocation-point budget \(N_r\) is independent of the reference-grid resolution, we tested whether fixing it could bias the scaling comparison. Varying \(N_r\) by factors of 2--6 in the 2D supersonic \(256^2\) case produces errors within the range of the benchmark suite, indicating that the fixed-budget choice does not qualitatively affect the scaling result (Appendix~\ref{app:collocation}).

Figure~\ref{fig:scaling} (right) quantifies the additional computational cost of increasing the spatial dimensionality. For each resolution, we compute the ratio $T_{\rm 3D}/T_{\rm 2D}$ for the subsonic runs. For Athena++, this ratio grows from ${\approx}6\times$ at $128$ to ${\approx}12\times$ at $1024$, reflecting the rapid growth in the number of grid cells and associated computational workload. For the PINN, the corresponding ratio remains nearly constant near $2\times$. Thus, although the present PINN implementation has substantially higher absolute training cost (see next paragraph), its computational scaling is markedly weaker than that of the grid-based solver. This difference becomes increasingly important as resolution and dimensionality grow, and motivates further optimization of PINN training for large-scale applications.

These comparisons concern scaling rather than absolute wall-clock efficiency: at the \(128^3\) 3D subsonic baseline, Athena++ requires approximately \(235\) s, whereas PINN training requires \(392\) min. Thus, the present implementation remains substantially slower in absolute terms, while exhibiting markedly weaker growth in computational cost with resolution and dimensionality.

\subsection{Robustness across turbulent initial conditions}
\label{sec:robustness}

The nonlinear evolution of a self-gravitating molecular cloud is highly sensitive to its turbulent initial conditions, with different realizations producing distinct morphologies including filaments, sheets, and compact collapsing cores. It is therefore important to assess whether the performance of the PINN depends strongly on the particular realization or remains consistent across a diverse set of turbulent flows. To this end, we evaluate the framework on twelve different configurations, with various combinations of random seed, mach number, and resolution.

\movetableright=-0.45in
\begin{table}
    \caption{Performance across different initializations}
    \label{tab:robustness_table}
    \centering
    \renewcommand{\arraystretch}{1.15}
    \begin{tabular}{lcccc}
    \hline\hline
    \multicolumn{5}{c}{\textbf{2D subsonic}} \\
    \hline
    Seed & Mach No. & Resolution & Mean $\delta_\rho$ & RMS $\delta_\rho$ \\
    \hline
    82 & \textbf{0.05} & 256 & 0.012 & 0.170 \\
    82 & \textbf{0.10} & 256 & 0.039 & 0.229 \\
    82 & \textbf{0.15} & 256 & 0.056 & 0.205 \\
    82 & \textbf{0.20} & 256 & 0.117 & 0.324 \\
    \hline
    \multicolumn{5}{c}{\textbf{2D supersonic}} \\
    \hline
    Seed & Mach No. & Resolution & Mean $\delta_\rho$ & RMS $\delta_\rho$ \\
    \hline
    82 & 1.1 & \textbf{128} & 0.098 & 0.235 \\
    82 & 1.1 & \textbf{256} & 0.045 & 0.146 \\
    82 & 1.1 & \textbf{512} & 0.042 & 0.135 \\
    82 & 1.1 & \textbf{1024} & 0.078 & 0.209 \\
    \hline
    \multicolumn{5}{c}{\textbf{3D subsonic}} \\
    \hline
    Seed & Mach No. & Resolution & Mean $\delta_\rho$ & RMS $\delta_\rho$ \\
    \hline
    \textbf{9} & 0.1 & 256 & 0.015 & 0.062 \\
    \textbf{38} & 0.1 & 256 & -0.004 & 0.068 \\
    \textbf{67} & 0.1 & 256 & -0.013 & 0.066 \\
    \textbf{82} & 0.1 & 256 & -0.026 & 0.062 \\
    \hline\hline
    \end{tabular}
    \tablecomments{$\delta_\rho = \log_2(\rho_{\rm PINN}/\rho_{\rm Athena++})$; mean and RMS are computed over the full spatial domain at the final output time.}
\end{table}

Table~\ref{tab:robustness_table} summarises the mean logarithmic bias and RMS density error across all twelve configurations. The mean bias remains small throughout with no systematic over- or under-estimation, and the RMS error stays within the same order regardless of Mach number, resolution, or dimensionality. These results confirm that the performance shown in the preceding subsections is representative of broadband turbulent self-gravitating flows rather than specific to a single realization.

\section{Discussion}

The results presented in this work demonstrate that physics-informed neural networks can reproduce the dominant evolution of turbulent, self-gravitating flows across a range of Mach numbers, resolutions, and spatial dimensions. The agreement with Athena++ across independent turbulent realizations indicates that the method is not restricted to a particular initial morphology. At the same time, the experiments reveal limitations as the flow becomes increasingly nonlinear and develops structure over a broad range of spatial and temporal scales. We discuss these limitations and possible directions for extending the work below.

\subsection{Limitations}
\label{sec:limitations}

\begin{figure*}
  \centering
  \includegraphics[width=0.995\textwidth]{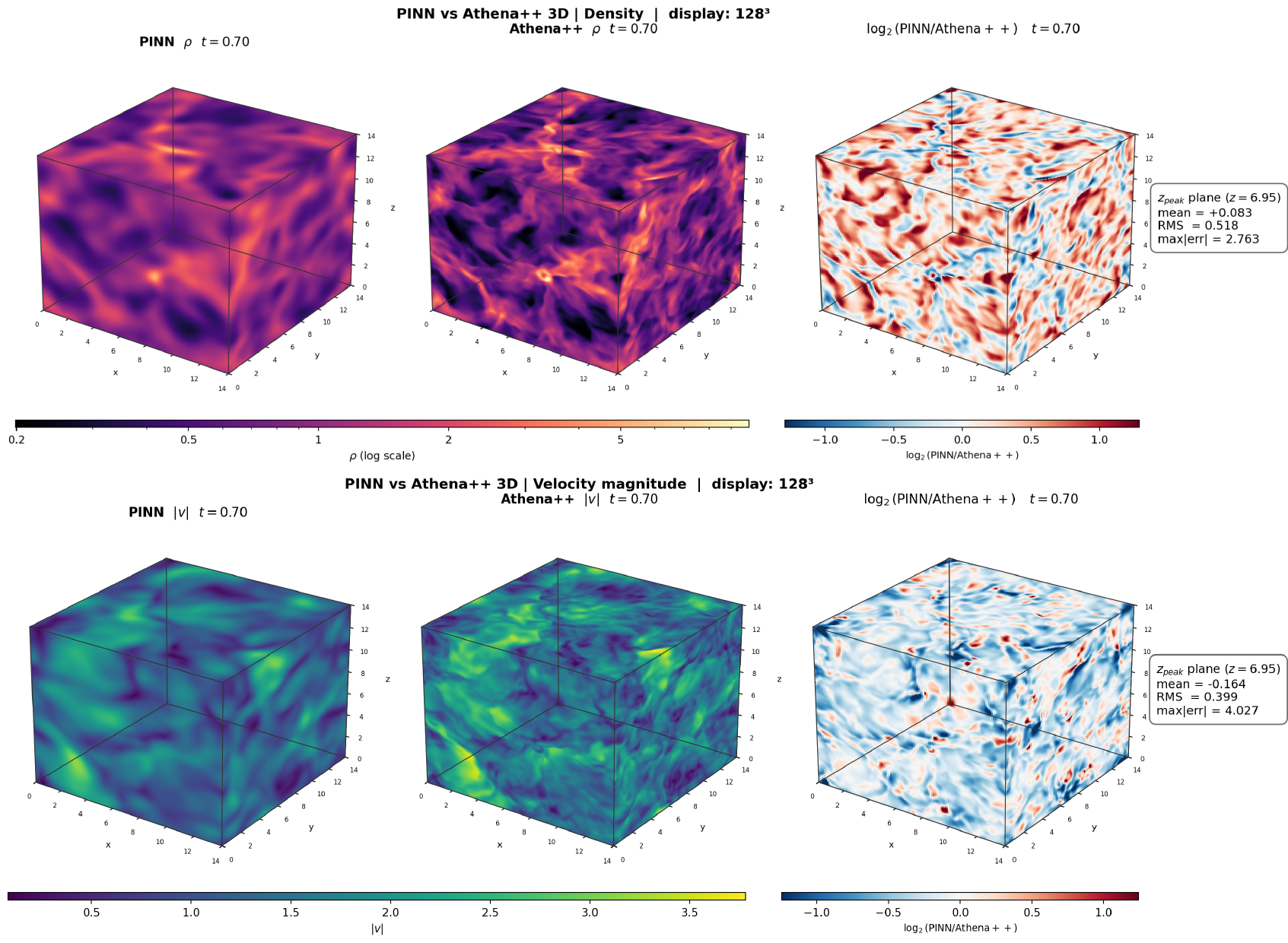}
  \caption{Same as Figure \ref{fig:3d_subsonic}, but with $a = 1.1$, representing a supersonic turbulence in the system}
  \label{fig:3d_supersonic}
\end{figure*}

Figure~\ref{fig:3d_supersonic} shows the PINN versus Athena++ comparison for the strongly supersonic three-dimensional case ($a = 1.1$), which provides a useful stress test of the present formulation. This case combines three-dimensional structure, supersonic compression, and rapidly developing gravitational collapse. At early times, the PINN captures the large-scale density and velocity organization seen in Athena++, reproducing the dominant flow morphology. Its accuracy, however, decreases as the flow develops increasingly dense and compact structures, with the density RMS log-ratio increasing to $0.518$ by $t=0.70$. The PINN therefore retains the large-scale character of the evolution but progressively loses fidelity in the most nonlinear stages of the collapse.

The smoother PINN solutions around sharp density gradients are consistent with the spectral bias of neural networks, which favours lower-frequency components over high-frequency structure \citep{wang2022and}, limiting the recovery of small-scale shocks and filamentary features as gravitational collapse generates progressively finer spatial structure.

A related difficulty is that PINN accuracy degrades as the solution becomes more rapidly evolving — a feature common to all four cases presented here rather than specific to the 3D supersonic run. As gravitational collapse proceeds and density growth rates increase, errors accumulate at late times regardless of Mach number or dimensionality; the onset simply occurs later in the subsonic cases where the dynamics accelerate more gradually. The 3D supersonic case represents the most extreme instance, with growth rates large enough that the PINN loses substantial fidelity within the training window.

Finally, the present implementation remains substantially more expensive in absolute wall-clock time than Athena++. The scaling experiments show much weaker growth of PINN training cost with spatial resolution and dimensionality, but the current implementation does not provide a computational speed advantage for the problem sizes considered here. The scaling result should therefore be interpreted as a potentially favourable scaling property rather than as a demonstration of computational superiority.

\subsection{Future Directions}

The limitations identified above suggest several directions for extending the current work to more strongly nonlinear and multi-scale flows. The degradation in the 3D supersonic stress test motivates architectures that can represent localized structure more effectively. Domain-decomposition approaches such as XPINNs and multiscale formulations such as FBPINNs provide natural possibilities, allowing different parts of the solution to be represented locally rather than relying on a single global network \citep{shukla2021parallel,moseley2023finite}. These approaches may be particularly relevant for separating the large-scale turbulent flow from localized shocked and collapsing structures.

The increasing difficulty at late times also motivates a more adaptive treatment of the space--time domain. Rather than representing the entire evolution with a single global network, future formulations could employ temporal decomposition \citep{mao2025neural} or progressively refined training as gravitational collapse accelerates. Adaptive collocation could provide a complementary mechanism for concentrating training effort in regions where the solution becomes difficult to represent, although its effectiveness in the strongly nonlinear regime remains to be established.

The computational cost of the present implementation provides a further motivation for distributed training. Multi-GPU and domain-decomposition strategies could reduce the training time required for larger collocation sets and higher-dimensional calculations, while retaining the physics-informed formulation \citep{shukla2021parallel}. Finally, extending beyond isothermal hydrodynamics to include non-isothermal thermodynamics, magnetic fields, radiation, and chemistry would provide a path toward increasingly realistic models of star-forming environments.

State-space-model-based architectures, such as PINNMamba \citep{xu2025sub}, offer another promising direction for improving the representation of spatiotemporal dependencies in PINNs.

\section{Summary and Conclusions}

We have demonstrated that physics-informed neural networks can model turbulent, self-gravitating molecular-cloud flows substantially beyond the controlled perturbations considered in GRINN. Using broadband turbulent velocity fields with a $k^{-4}$ power spectrum, we tested isothermal self-gravitating gas across subsonic and supersonic regimes, two- and three-dimensional geometries, multiple resolutions, and independent turbulent realizations. The PINN consistently reproduced the dominant density and velocity evolution obtained with Athena++, with characteristic density discrepancies of approximately $5$--$17\%$ for the principal benchmark cases. The method also retained comparable performance across independent turbulent realizations, indicating that its behaviour is not tied to a particular initial morphology.

The calculations also reveal an important distinction between the scaling and the absolute computational cost of the present approach. Although PINN training is substantially more expensive in wall-clock time for the problem sizes considered here, its training cost increases by less than $25\%$ between the lowest and highest resolutions tested, in contrast to the much steeper scaling of the grid-based calculations. This suggests that the mesh-free formulation offers favourable scaling properties as problem resolution and dimensionality increase, while emphasizing that substantial optimization is still required before PINNs become computationally competitive with established grid-based methods.

The strongly supersonic three-dimensional stress test further identifies a clear limitation of the present formulation. As shocks and gravitational collapse produce increasingly dense, fine-scale, and rapidly evolving structure, the PINN progressively loses fidelity relative to Athena++. This result highlights the difficulty of representing strongly nonlinear, multi-scale evolution over extended time intervals and provides a concrete target for future methodological improvements.

Overall, these results establish PINNs as a complementary framework for modeling turbulent self-gravitating flows and extend physics-informed modeling substantially beyond previously studied controlled perturbations. Future progress will require improved representations of multi-scale structure, more effective training strategies, and extensions of the physical model toward the richer physics of realistic star-forming environments.

AI tools were used to assist with manuscript editing and refinement, as well as with development of the related code. ChatGPT, Claude, and Gemini were used for manuscript-related assistance; Cursor and Antigravity were used primarily for code development and documentation. All AI-assisted content and code used in the work were reviewed by the authors, who take full responsibility for the accuracy and integrity of the manuscript.

The code used in this work is publicly available on GitHub at \url{https://github.com/Tirthankar4/CC-PINN} and archived on Zenodo at \citep{tirthankar_de_2026_23294396}.

\begin{acknowledgements}
This research was enabled in part by support provided by Compute Ontario and the Digital Research Alliance of Canada. S. B. was supported by a Discovery Grant from the National Sciences and Engineering Research Council of Canada.
\end{acknowledgements}

\software{
Athena++ \citep{Stone2020},
PyTorch \citep{paszke2019pytorch},
NumPy \citep{harris2020array},
SciPy \citep{virtanen2020scipy},
Matplotlib \citep{hunter2007matplotlib},
h5py 3.16.0 \url{https://www.h5py.org/}
}

\bibliographystyle{aasjournalv7}
\bibliography{references}

\begin{thebibliography}{}
\expandafter\ifx\csname natexlab\endcsname\relax\def\natexlab#1{#1}\fi
\providecommand{\url}[1]{\href{#1}{#1}}
\providecommand{\dodoi}[1]{doi:~\href{http://doi.org/#1}{\nolinkurl{#1}}}
\providecommand{\doeprint}[1]{\href{http://ascl.net/#1}{\nolinkurl{http://ascl.net/#1}}}
\providecommand{\doarXiv}[1]{\href{https://arxiv.org/abs/#1}{\nolinkurl{https://arxiv.org/abs/#1}}}

\bibitem[{S. Auddy {et~al.}(2018)Auddy, Basu, \& Kudoh}]{auddy2018effect}
Auddy, S., Basu, S., \& Kudoh, T. 2018, \bibinfo{title}{The effect of magnetic fields and ambipolar diffusion on the column density probability distribution function in molecular clouds,} Monthly Notices of the Royal Astronomical Society, 474, 400

\bibitem[{S. Auddy {et~al.}(2024)Auddy, Dey, Turner, \& Basu}]{auddy2024grinn}
Auddy, S., Dey, R., Turner, N.~J., \& Basu, S. 2024, \bibinfo{title}{GRINN: a physics-informed neural network for solving hydrodynamic systems in the presence of self-gravity,} Machine Learning: Science and Technology, 5, 025014

\bibitem[{R. Banerjee {et~al.}(2004)Banerjee, Pudritz, \& Holmes}]{banerjee2004formation}
Banerjee, R., Pudritz, R.~E., \& Holmes, L. 2004, \bibinfo{title}{The formation and evolution of protostellar discs; three-dimensional adaptive mesh refinement hydrosimulations of collapsing, rotating Bonnor--Ebert spheres,} Monthly Notices of the Royal Astronomical Society, 355, 248

\bibitem[{M. Chevance {et~al.}(2020)Chevance, Kruijssen, Vazquez-Semadeni, Nakamura, Klessen, Ballesteros-Paredes, Inutsuka, Adamo, \& Hennebelle}]{chevance2020molecular}
Chevance, M., Kruijssen, J.~D., Vazquez-Semadeni, E., {et~al.} 2020, \bibinfo{title}{The molecular cloud lifecycle,} Space Science Reviews, 216, 50

\bibitem[{T. De(2026)De}]{tirthankar_de_2026_23294396}
De, T. 2026, Tirthankar4/CC-PINN: v1.0.0, v1.0.0 Zenodo, \dodoi{10.5281/zenodo.23294396}

\bibitem[{C. Federrath {et~al.}(2010)Federrath, Roman-Duval, Klessen, Schmidt, \& Mac~Low}]{federrath2010comparing}
Federrath, C., Roman-Duval, J., Klessen, R.~S., Schmidt, W., \& Mac~Low, M.-M. 2010, \bibinfo{title}{Comparing the statistics of interstellar turbulence in simulations and observations: solenoidal versus compressive turbulence forcing,} Astronomy \& Astrophysics, 512, A81

\bibitem[{C.~R. Harris {et~al.}(2020)Harris, Millman, Van Der~Walt, Gommers, Virtanen, Cournapeau, Wieser, Taylor, Berg, Smith, {et~al.}}]{harris2020array}
Harris, C.~R., Millman, K.~J., Van Der~Walt, S.~J., {et~al.} 2020, \bibinfo{title}{Array programming with NumPy,} nature, 585, 357

\bibitem[{M.~H. {Heyer} \& C.~M. {Brunt}(2004){Heyer} \& {Brunt}}]{heyer04}
{Heyer}, M.~H., \& {Brunt}, C.~M. 2004, \bibinfo{title}{{The Universality of Turbulence in Galactic Molecular Clouds},} \apjl, 615, L45, \dodoi{10.1086/425978}

\bibitem[{J.~D. Hunter(2007)Hunter}]{hunter2007matplotlib}
Hunter, J.~D. 2007, \bibinfo{title}{Matplotlib: A 2D graphics environment,} Computing in science \& engineering, 9, 90

\bibitem[{D.~C. Liu \& J. Nocedal(1989)Liu \& Nocedal}]{liu1989limited}
Liu, D.~C., \& Nocedal, J. 1989, \bibinfo{title}{On the limited memory BFGS method for large scale optimization,} Mathematical programming, 45, 503

\bibitem[{M.-M. Mac~Low \& R.~S. Klessen(2004)Mac~Low \& Klessen}]{mac2004control}
Mac~Low, M.-M., \& Klessen, R.~S. 2004, \bibinfo{title}{Control of star formation by supersonic turbulence,} Reviews of modern physics, 76, 125

\bibitem[{S. Mao {et~al.}(2025)Mao, Wang, Wang, Dong, Lu, Yi, Perdikaris, Isella, Fabbro, \& Wang}]{mao2025neural}
Mao, S., Wang, W., Wang, S., {et~al.} 2025, \bibinfo{title}{Neural Networks as Surrogate Solvers for Time-dependent Accretion Disk Dynamics,} The Astrophysical Journal Letters, 992, L20

\bibitem[{C.~F. {McKee} \& E.~C. {Ostriker}(2007){McKee} \& {Ostriker}}]{mckee07}
{McKee}, C.~F., \& {Ostriker}, E.~C. 2007, \bibinfo{title}{{Theory of Star Formation},} \araa, 45, 565, \dodoi{10.1146/annurev.astro.45.051806.110602}

\bibitem[{B. Moseley {et~al.}(2023)Moseley, Markham, \& Nissen-Meyer}]{moseley2023finite}
Moseley, B., Markham, A., \& Nissen-Meyer, T. 2023, \bibinfo{title}{Finite basis physics-informed neural networks (FBPINNs): a scalable domain decomposition approach for solving differential equations: B. Moseley et al.,} Advances in Computational Mathematics, 49, 62

\bibitem[{G. Parascandolo {et~al.}(2017)Parascandolo, Huttunen, \& Virtanen}]{parascandolo2017taming}
Parascandolo, G., Huttunen, H., \& Virtanen, T. 2017, Taming the waves: sine as activation function in deep neural networks, \url{https://openreview.net/forum?id=Sks3zF9eg}

\bibitem[{A. Paszke {et~al.}(2019)Paszke, Gross, Massa, Lerer, Bradbury, Chanan, Killeen, Lin, Gimelshein, Antiga, {et~al.}}]{paszke2019pytorch}
Paszke, A., Gross, S., Massa, F., {et~al.} 2019, \bibinfo{title}{Pytorch: An imperative style, high-performance deep learning library,} Advances in neural information processing systems, 32

\bibitem[{M. Raissi {et~al.}(2019)Raissi, Perdikaris, \& Karniadakis}]{raissi2019physics}
Raissi, M., Perdikaris, P., \& Karniadakis, G.~E. 2019, \bibinfo{title}{Physics-informed neural networks: A deep learning framework for solving forward and inverse problems involving nonlinear partial differential equations,} Journal of Computational physics, 378, 686

\bibitem[{K. Shukla {et~al.}(2021)Shukla, Jagtap, \& Karniadakis}]{shukla2021parallel}
Shukla, K., Jagtap, A.~D., \& Karniadakis, G.~E. 2021, \bibinfo{title}{Parallel physics-informed neural networks via domain decomposition,} Journal of Computational Physics, 447, 110683

\bibitem[{J.~M. Stone {et~al.}(2020)Stone, Tomida, White, \& Felker}]{Stone2020}
Stone, J.~M., Tomida, K., White, C.~J., \& Felker, K.~G. 2020, \bibinfo{title}{The Athena$\mathplus$$\mathplus$ Adaptive Mesh Refinement Framework: Design and Magnetohydrodynamic Solvers,} The Astrophysical Journal Supplement Series, 249, 4, \dodoi{10.3847/1538-4365/ab929b}

\bibitem[{P. Virtanen {et~al.}(2020)Virtanen, Gommers, Oliphant, Haberland, Reddy, Cournapeau, Burovski, Peterson, Weckesser, Bright, {et~al.}}]{virtanen2020scipy}
Virtanen, P., Gommers, R., Oliphant, T.~E., {et~al.} 2020, \bibinfo{title}{SciPy 1.0: fundamental algorithms for scientific computing in Python,} Nature methods, 17, 261

\bibitem[{S. Wang {et~al.}(2022)Wang, Yu, \& Perdikaris}]{wang2022and}
Wang, S., Yu, X., \& Perdikaris, P. 2022, \bibinfo{title}{When and why PINNs fail to train: A neural tangent kernel perspective,} Journal of Computational Physics, 449, 110768

\bibitem[{R.~E. Wengert(1964)Wengert}]{wengert1964simple}
Wengert, R.~E. 1964, \bibinfo{title}{A simple automatic derivative evaluation program,} Communications of the ACM, 7, 463

\bibitem[{C. Xu {et~al.}(2025)Xu, Liu, Hu, Li, Qin, Zheng, \& Xiong}]{xu2025sub}
Xu, C., Liu, D., Hu, Y., {et~al.} 2025, \bibinfo{title}{Sub-sequential physics-informed learning with state space model,} arXiv preprint arXiv:2502.00318

\bibitem[{Z. Zhang(2018)Zhang}]{zhang2018improved}
Zhang, Z. 2018, \bibinfo{title}{Improved adam optimizer for deep neural networks,} in 2018 IEEE/ACM 26th international symposium on quality of service (IWQoS), Ieee, 1--2

\end{thebibliography}

\appendix
\renewcommand*{\theHequation}{appendix.\thesection.\arabic{equation}}
\vspace*{-18pt}

\section{PINN Implementation Details} \label{app:pinn}

\subsection{Turbulent Initial-Condition Field Generation}
\label{app:icgen}

\paragraph{Synthesis algorithm.}
Each Cartesian velocity component $v_i$ ($i = x, y$ in 2-D; $i = x, y, z$ in 3-D) is synthesised independently from a separate white-noise draw using the following pipeline:
\begin{enumerate}
  \item A real-valued white-noise field $\xi(\bm{x})$ is drawn from a standard
        normal distribution on a regular Cartesian grid covering the periodic domain
        $[0, L]^d$.
  \item The field is transformed to Fourier space via a real fast Fourier transform,
        giving $\tilde{\xi}(\bm{k})$ with Hermitian symmetry by construction, so
        that the inverse transform is purely real.
  \item A power-law filter is applied in Fourier space,
        \begin{equation}
          \tilde{v}_i(\bm{k}) =
          \begin{cases}
            k^{-2}\,\tilde{\xi}(\bm{k}) & k > 0, \\
            0 & k = 0,
          \end{cases}
          \label{eq:ic_filter}
        \end{equation}
        where $k = |\bm{k}|$ is the isotropic wavenumber. Setting the
        $\bm{k}=\bm{0}$ mode to zero enforces zero spatial mean and hence
        zero net momentum in each component.
  \item The filtered spectrum is transformed back to physical space via an inverse
        real FFT, yielding a real field $v_i(\bm{x})$ with power spectrum
        $P_{v_i}(k) \propto k^{-4}$.
  \item The field is rescaled so that $\mathrm{std}(v_i) = a\,c_s$ exactly,
        matching the target per-component RMS Mach number.
\end{enumerate}
Because each component is drawn from an independent white-noise realisation, the resulting velocity field is statistically isotropic in expectation, with no preferred direction imposed by the synthesis procedure.

\paragraph{Shared-field protocol.}
To guarantee that the PINN and Athena++ start from identical initial conditions, the velocity arrays are generated \emph{once} on a regular Cartesian grid and shared between both solvers. For Athena++, the grid values are assigned directly to the finite-volume cell centres. For the PINN, the grid values are bilinearly interpolated to arbitrary collocation-point coordinates using a \texttt{RegularGridInterpolator} \citep{virtanen2020scipy} at training time, whenever the IC loss $\mathcal{L}_{\mathrm{IC}}$ (Appendix~\ref{app:loss}) is evaluated. The IC generation grid is matched to the Athena++ resolution for each experiment, so the interpolation to PINN collocation points introduces negligible error relative to the finite-volume truncation error.

\paragraph{Reproducibility.}
The entire field is determined by a single integer random seed passed to NumPy's default random generator \citep{harris2020array}. For each experiment, the random seed is held fixed when generating the shared realization used by both solvers.

\subsection{Periodic Feature Embedding} \label{app:fourier}

Rather than feeding raw spatial coordinates into the network, each spatial coordinate $u \in [u_{\min}, u_{\max}]$ is first mapped to a set of periodic features. For harmonic index $k = 1, \ldots, H$, the embedding is

\begin{equation}
  \bm{\phi}_k(u)
    = \frac{1}{\sqrt{k}}
      \begin{bmatrix}
        \sin(k\,\theta) \\[2pt]
        \cos(k\,\theta)
      \end{bmatrix},
  \qquad
  \theta = \frac{2\pi\,(u - u_{\min})}{L},
\end{equation}
where $L = u_{\max} - u_{\min}$ is the domain length and $H$ is the number of harmonics.  The time coordinate is passed through unchanged.  The periodic sine/cosine basis is exactly consistent with the periodic boundary conditions imposed on the domain, satisfying them by construction without penalty terms in the loss. The $1/\sqrt{k}$ factor rescales the amplitude of higher harmonics at the network input, mildly reducing the scale of higher harmonics at the network input by a factor of about 1.7 between $k=1$ and $k=3$. It is a heuristic conditioning choice rather than a constraint: it reduces the initial contribution of high-wavenumber features to the first-layer pre-activations and slows their effective learning rate, biasing early training toward smooth, low-wavenumber solutions.

\subsection{Exponential Density Parameterisation} \label{app:exparam}

The network does not output physical density $\rho$ directly. Instead, it outputs a raw field $\hat{\rho}$, and the physical density is recovered via

\begin{equation}
  \rho(\bm{x}, t)
    = \rho_0\,\exp\!\bigl(\mathrm{clamp}(t_{\mathrm{eff}}\,\hat{\rho},\;
      -4,\; 4)\bigr),
  \qquad
  t_{\mathrm{eff}} = \max(t - \Delta t_s,\; 0),
\end{equation}
where $\Delta t_s$ is the startup delay described in Appendix~\ref{app:startup}.  This parameterisation enforces two physical constraints as hard constraints, independent of the network weights:

\begin{enumerate}
  \item \textbf{Strict positivity.}  Because $\exp(\cdot) > 0$ for all real
        arguments, the density is guaranteed positive for any network output,
        eliminating the need for a positivity penalty in the loss.

  \item \textbf{Uniform-density initial condition.}  At $t = 0$,
        $t_{\mathrm{eff}} = 0$ and $\rho(\bm{x}, 0) = \rho_0\exp(0) =
        \rho_0$ exactly, regardless of $\hat{\rho}$.
\end{enumerate}

The exponential form is also convenient for representing multiplicative density growth: the network effectively learns the logarithm of the density relative to its initial value. The clamp bounds $-4,4$ prevent loss blowup at initialization, when the product $t_{\mathrm{eff}}\hat{\rho}$ can be large before training has converged; they correspond to density ratios of $e^{-4}\approx0.018$ and $e^4\approx55$ relative to $\rho_0$, safely bracketing the range of densities encountered in the simulations.

\subsection{PDE Enforcement Startup Delay} \label{app:startup}

We introduce a fixed startup delay  $\Delta t_s = 0.1t_0$, during which the PDE residuals are not enforced. This is implemented in two complementary ways.

\paragraph{Collocation point sampling.}
Residual collocation points are drawn uniformly from the time interval $[\Delta t_s,\, t_{\max}]$.  No PDE residual is evaluated at any $t < \Delta t_s$, so the network develops its internal representation of the flow during this interval without being penalised against the governing equations.

\paragraph{Density mask.}
The effective time $t_{\mathrm{eff}} = \max(t - \Delta t_s, 0)$ in the exponential parametrisation (Appendix~\ref{app:exparam}) evaluates to zero for $t < \Delta t_s$, freezing the density exactly at $\rho_0$ as a hard constraint.  For $t \geq \Delta t_s$, the density evolves freely according to the learned growth-rate field.

This treatment is motivated by the causal development of the initial velocity perturbation. The imposed velocity field propagates and drives the subsequent density response at finite characteristic speeds. Allowing a short interval before enforcing the fully coupled PDE system therefore provides a controlled way for the initially imposed velocity field to begin
developing before density evolution is constrained by the PDE residuals. The startup delay is a training strategy introduced for the present problem, rather than a physical modification of the hydrodynamic evolution.

\subsection{Loss Function} \label{app:loss}

The general PINN loss formulation, including the explicit forms of the PDE residuals for the continuity, momentum, and Poisson equations, follows \citet{auddy2024grinn}, to which we refer the reader for a detailed derivation. Here we describe only the modifications introduced in the present work.

\paragraph{Boundary-condition loss: eliminated by hard constraint.} In \citet{auddy2024grinn}, periodic boundary conditions were enforced as a soft penalty term ($MSE_b$) in the training loss.  In the present work this term is absent entirely: the periodic feature embedding (Appendix~\ref{app:fourier}) maps each spatial coordinate onto a periodic sine/cosine basis, so the resulting network representation is periodic by construction and no explicit boundary-condition penalty is required. The total training loss therefore reduces to
\begin{equation}
  \mathcal{L} = \mathcal{L}_{\mathrm{IC}} + \mathcal{L}_{\mathrm{PDE}},
\end{equation}
compared with $MSE_{\mathrm{PDE}} + MSE_b + MSE_0$ in \citet{auddy2024grinn}.

\paragraph{Initial-condition loss.}
The density initial condition $\rho(\bm{x},0)=\rho_0$ is enforced as a hard constraint via the exponential parametrisation (Appendix~\ref{app:exparam}), contributing no loss term.  The velocity initial conditions, which now encode a broadband turbulent field rather than a sinusoidal mode, are enforced as a soft constraint,
\begin{equation}
  \mathcal{L}_{\mathrm{IC}}
    = \frac{1}{N_0} \sum_{j=1}^{N_0}
      \sum_{i=1}^{d}
      \bigl[v_i(\bm{x}^{(j)}, 0) - v_{i,\mathrm{turb}}(\bm{x}^{(j)})\bigr]^2,
\end{equation}
where $N_0$ is the number of IC collocation points and $v_{i,\mathrm{turb}}$ is the prescribed turbulent velocity field.

\paragraph{PDE residual loss.}
The PDE residuals of Eqs.~(\ref{eq:continuity})--(\ref{eq:poisson}) are evaluated at $N_r$ collocation points sampled from $[\Delta t_s, t_{\max}]$ (Appendix~\ref{app:startup}) via automatic differentiation,
\begin{equation}
  \mathcal{L}_{\mathrm{PDE}}
    = \frac{1}{N_r}\sum_{j=1}^{N_r}
      \Bigl[
        r_\rho^2
        + \sum_{i=1}^{d} r_{v_i}^2
        + r_\phi^2
      \Bigr]^{(j)},
\end{equation}
where $r_\rho$, $r_{v_i}$, and $r_\phi$ are the pointwise residuals of the continuity, momentum, and Poisson equations respectively (see \citealt{auddy2024grinn}, Eqs.~6--8, for their explicit forms).

During Adam training, gradients are clipped to unit norm before each parameter update to prevent gradient explosions early in training.  During L-BFGS, the same collocation batch is cached and reused across all line-search function evaluations within a single optimizer step, ensuring a consistent loss landscape required for convergence.

\subsection{Network Hyperparameters} \label{app:hyperparams}

Table~\ref{tab:hyperparams} summarises the hyperparameters used for all runs reported in this paper. The same architecture and optimizer settings were applied uniformly across all cases, with no tuning to individual Mach numbers, spatial resolutions, or turbulent realizations. The consistent performance of PINN across the full benchmark suite (Section~\ref{sec:robustness}) therefore reflects the behaviour of a fixed model configuration rather than one optimized separately for each case.

\begin{table}[h]
  \caption{PINN hyperparameters used in all experiments.}
  \label{tab:hyperparams}
  \centering
  \renewcommand{\arraystretch}{1.2}
  \begin{tabular}{lcc}
    \hline\hline
    Parameter & Symbol & Value \\
    \hline
    \multicolumn{3}{c}{\textit{Architecture}} \\
    \hline
    Hidden layers              & $N_\ell$              & 7 \\
    Neurons per layer          & $N_h$                 & 128 \\
    Harmonics          & $H$                   & 3 \\
    Density parameterisation   & ---                   & Exponential \\
    \hline
    \multicolumn{3}{c}{\textit{Collocation points}} \\
    \hline
    IC points                  & $N_0$                 & 130\,000 \\
    PDE residual points        & $N_r$                 & 780\,000 \\
    Batch size                 & ---                   & 130\,000 \\
    \hline
    \multicolumn{3}{c}{\textit{Optimisation}} \\
    \hline
    Adam iterations            & ---                   & 6\,001 \\
    Adam learning rate         & $\eta$                & $10^{-3}$ \\
    Adam gradient clip norm    & ---                   & 1.0 \\
    L-BFGS iterations          & ---                   & 251 \\
    L-BFGS line search         & ---                   & Strong Wolfe \\
    Startup delay              & $\Delta t_s$          & 0.1 \\
    \hline\hline
  \end{tabular}
\end{table}

\section{Collocation-Point Budget}
\label{app:collocation}

To verify that the near-constant PINN training cost across spatial resolutions is not an artefact of holding the collocation-point budget $N_r$ fixed, we ran the 2D supersonic $256^2$ case (seed 82) at reduced budgets while keeping all other hyperparameters identical to Table~\ref{tab:hyperparams}. The baseline uses $N_r=780{,}000$ collocation points, while the reduced-budget runs use $N_r=390{,}000$ and $130{,}000$.

The resulting mean and RMS density errors $(\langle\delta_\rho\rangle,\mathrm{RMS}[\delta_\rho])$ are $(0.106,0.266)$, $(0.106,0.286)$, and $(0.045,0.146)$ for $N_r=130{,}000$, $390{,}000$, and $780{,}000$, respectively. The errors remain within the range found across the benchmark suite, with no systematic monotonic degradation as $N_r$ is reduced. The lower RMS error at the baseline budget indicates some sensitivity to $N_r$ in this realization, but the results provide no evidence that fixing the collocation budget qualitatively biases the resolution-scaling comparison. We therefore retain a fixed $N_r$ across resolutions.

\begin{figure*}
  \centering
  \includegraphics[width=0.85\textwidth]{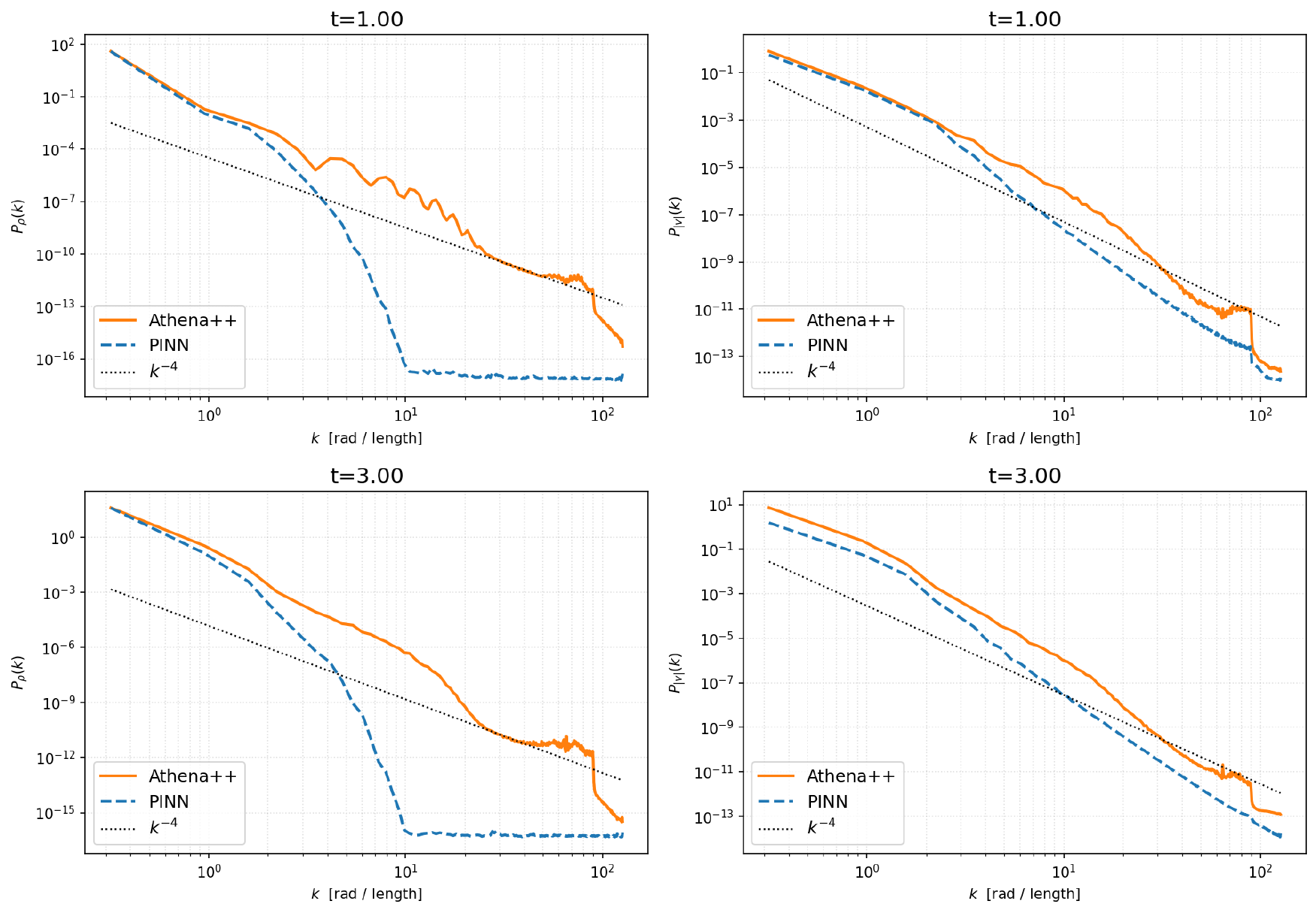}
  \caption{Radially averaged power spectra of the density field $P_\rho(k)$
    (left) and velocity magnitude $P_{|\bm{v}|}(k)$ (right) for the 2D
    subsonic baseline case at $t \simeq 1.0\,t_0$ (top) and $t = 3.00\,t_0$
    (bottom). Solid orange: Athena++; dashed blue: PINN; dotted black:
    $k^{-4}$ reference slope. The PINN and Athena++ solutions show similar
    large-scale spectral behaviour, while the PINN increasingly underestimates
    high-$k$ density power as gravitational collapse develops finer-scale
    structure. The velocity-magnitude spectra remain in closer agreement over
    a broader range of scales.}
    \label{fig:power_spectrum}
\end{figure*}

\section{Power Spectrum Comparison}
\label{app:power_spectrum}

Figure~\ref{fig:power_spectrum} shows the radially averaged power spectra of the density field $P_\rho(k)$ and velocity magnitude $P_{|\bm{v}|}(k)$ for the 2D subsonic baseline case at $t \simeq 1.0\,t_0$ and $t = 3.00\,t_0$, comparing the PINN and Athena++ solutions.

At large scales, the PINN and Athena++ spectra show similar overall behavior, with the velocity-magnitude spectra remaining particularly close in both shape and slope. The dotted line indicates a $k^{-4}$ reference slope corresponding to the spectral form used to initialize the turbulent velocity field; it is not an expected slope for the evolved density
spectrum. The density power increases substantially as gravitational collapse develops, while the PINN retains the same overall large-scale evolution but with reduced power.

The main discrepancy appears at higher wavenumbers, where the PINN underestimates density power relative to Athena++. This deficit is more pronounced at later times, when gravitational collapse produces increasingly fine-scale density structure. The result is consistent with the tendency of the PINN solution toward smoother, lower-frequency structure discussed in Section~\ref{sec:limitations}, and with the finite spatial bandwidth of the $H=3$ harmonic embedding (Appendix~\ref{app:fourier}). In contrast, the velocity-magnitude spectra remain in closer agreement over a broader range of wavenumbers, indicating that the loss of small-scale spectral power is more pronounced for density than for velocity over the simulated interval.

\end{document}